\documentclass[twocolumn,aps,showpacs,amsmath,amssymb]{revtex4-2}
\usepackage{amsmath,amssymb}
\usepackage{citesort}
\usepackage{graphicx}
\begin{document}

\title{Microscopic strain correlations in sheared amorphous solids}
\author{Sagar Malik$^{1}$, Meenakshi L.$^{2}$, Atharva Pandit$^{1}$, Antina Ghosh$^{1}$, Peter Schall$^{3}$, Bhaskar Sengupta$^{2}$, Vijayakumar Chikkadi$^1$}
\affiliation{
$^1$ Physics Division, Indian Institute of Science Education and Research Pune, 411008 Pune, India.\\
$^2$ Department of Physics, School of Advanced Sciences, Vellore Institute of Technology, 632014 Vellore, India.\\
$^3$ Institute of Physics, University of Amsterdam, 1098 XH Amsterdam, The Netherlands.\\
}

\begin{abstract}
We investigate spatial correlations of strain fluctuations in sheared colloidal glasses and simulations of sheared amorphous solids. The correlations reveal a quadrupolar symmetry reminiscent of the strain field due to an Eshelby's inclusion. However, they display an algebraic decay $1/r^{\alpha}$, where the exponent $\alpha$ is close to $1$ in the steady state, unlike the Eshelby field, for which $\alpha=3$ . The exponent takes values between $3$ to $1$ in the transient stages of deformation. We explain these observations using a simple model based on interacting Eshelby inclusions. As the system is sheared beyond the linear response to plastic flow, the density correlations of inclusions are enhanced and it emerges as key to understanding the elastoplastic response of the system to applied shear.  
\end{abstract}

\maketitle

Amorphous solids are an important class of materials that appear in various forms ranging from metallic glasses to polymeric glasses and soft materials made of emulsions, foams and granular matter \cite{Barrat18, Biroli11}. Even though their mechanical properties differ significantly, they display similar elastic and plastic properties. Therefore, the elastoplastic deformation of amorphous solids has attracted considerable attention in the past decade, and it has been a topic of intense research \cite{Barrat18, Barrat11}. Due to disordered structure, the dislocation-based models of crystal plasticity cannot be extended to amorphous plasticity \cite{Barrat18, Falk98}. So, several questions relating to plasticity carriers in amorphous solids, such as their characteristics, formation, and spatiotemporal organization, are central to our understanding of this topic.
  
In recent years, simulations and experiments have established that plastic deformation in amorphous solids occurs due to localized rearrangement of particles associated with long-range quadrupolar strain field \cite{Lemaitre04, Lemaitre06, Barrat06, Schall07, Chikkadi11}. The strain field resembles the Eshelby elastic field around an inclusion in a homogeneous isotropic solid \cite{Eshelby57, Eshelby59}. The overall deformation occurs due to spatiotemporal interactions of such plastic rearrangements mediated by elasticity. Some of these basic features of amorphous plasticity have been an integral part of early elastoplastic models \cite{Spaepen77, Argon79, Hebraud98, Bocquet04}. Recent theoretical models have exploited these ideas to explain the formation of shear bands and yielding in quasi-static MD simulations of sheared amorphous solids \cite{Ratul12, Ratul13-1, Zaccone17}. Further, the mesoscopic simulations of amorphous plasticity based on lattice models have also incorporated the long-range nature of elastic strain fields arising from plastic rearrangements \cite{Dahmen09, Zoe17, Martens21}. The experimental investigations along these lines are scarce \cite{Chikkadi11, Schall07}. The previous experiments on sheared colloidal glasses had investigated spatial correlations of strain, and non-affine displacements in the steady state \cite{Chikkadi11, Chikkadi12}. Surprisingly, the spatial correlations of non-affine displacements were found to decay as $1/r$, instead of $1/r^2$ as predicted by Eshelby's solution. The anomalous exponent was proposed to originate from the interactions of several inclusions \cite{Chikkadi15}. However, there is no experimental evidence; rather, the nature of these correlations in the transient stages of deformation remains unclear. What is known is that the spatial correlations of strain in quiescent colloidal glasses or the linear response regime decay as $1/r^3$ \cite{Varnik18}. Therefore, the nature of the strain correlations as the system is sheared beyond the linear response to plastic flow is unknown. It is natural to ask whether a simple model based on interacting Eshelby-like inclusions can offer new insights into microscopic strain fluctuations and their correlations in the transient and steady states of deformation of amorphous solids.    

This paper presents a combined experimental and numerical investigation of microscopic strain fluctuations and their spatial correlations in 3D amorphous solids under shear. The experiments are performed using dense colloidal suspensions, and the simulations are done using dense binary mixture at low temperatures following molecular dynamics technique. The spatial correlations of strain fluctuations reveal a familiar quadrupolar symmetry; however, the correlations display an anomalous decay of $1/r^{\alpha}$, where the exponent $\alpha\sim1$ in the steady state and it varies from $\alpha\sim3-1$ in the transient stages of deformation. We provide an explanation for these observations using ideas that are motivated by mesoscopic models of amorphous plasticity. We identify inclusion centers based on the principal component of shear strain $\epsilon_{xz}$. All the particles with $\epsilon_{xz}$ larger than a threshold are considered centers of Eshelby inclusions. Further, the inclusion centers are used to determine synthetic strain maps following the simple superposition principle. The spatial correlations of synthetic strains reveal the origin of the anomalous exponents $\alpha$ and further highlight the role of spatial clustering of inclusions on the elastoplastic deformation of amorphous solids.     

We prepare a colloidal glass by suspending sterically stabilized fluorescent polymethylmethacrylate particles in a density and refractive index matching mixture of cycloheptyl bromide and cis-decalin. The particles have a diameter of $\sigma = 1.3 \mu m$, and a polydisperity of $7\%$ to prevent crystallization. The suspension of a desired volume fraction $\phi\sim0.60$ is prepared via centrifugation, and it is sheared using a shear cell that has two parallel boundaries \cite{Chikkadi11}. The $3D$ imaging of the particles during shear is done using a confocal microscope. The shear rate in our experiment is $\dot{\gamma}\sim1.5\times 10^{-5}s^{-1}$. Further details are included in the experimental section of supplementary information. In our molecular dynamics simulations, we prepare glass samples by initially equilibrating the binary mixture in a liquid state at temperature $T=1.0$ and gradually cooling it to a final temperature of $T=0.3$ at a slow cooling rate of $10^{-5}$ to obtain well relaxed samples. The simulations are done at a density of $\rho=1.2$, a system size of $N=150000$ particles, and a time step of $\Delta t =0.005$ with boundary conditions in all directions. The glassy samples are sheared along the $xz$ plane at a constant shear rate $\dot{\gamma}=10^{-4}$. The statistical averaging is done using $100$ realization with different initial conditions. The other simulation details are provided in the simulations section of supplementary information. 

The particle coordinates are used to compute microscopic strain and their spatial correlations. Briefly, to compute the particle level strain, the displacement of a particle relative to its first neighbors are considered to define a best fit affine strain tensor. The best fit tensor minimizes the non-affine displacement for the particle \cite{Falk98}. The principal shear strain component $\epsilon_{xz}$ is obtained from the strain tensor. We will investigate the correlations in the fluctuations of $\epsilon_{xz}$ using the following expression \cite{Chikkadi11}:
\begin{equation}
C_{\epsilon}({\bf \Delta r}) = \frac{ \left< \epsilon_{xz}({\bf r + \Delta r}) \epsilon_{xz}({\bf r}) \right> - \left< \epsilon_{xz}({\bf r}) \right> ^{2} } { \left< \epsilon_{xz}({\bf r})^{2} \right> - \left< \epsilon_{xz}({\bf r}) \right> ^{2} }  ,
\label{c_r}
\end{equation}
where angular brackets denote average over all the particles in the systems and several strain steps. $C_{\epsilon}$ correlates values of $\epsilon_{xz}$ at locations separated by ${\bf \Delta r}$, this way we capture the elasto-plastic response of the system in the steady state.

\begin{figure}[htp]
\centering
\includegraphics[width=.4\textwidth]{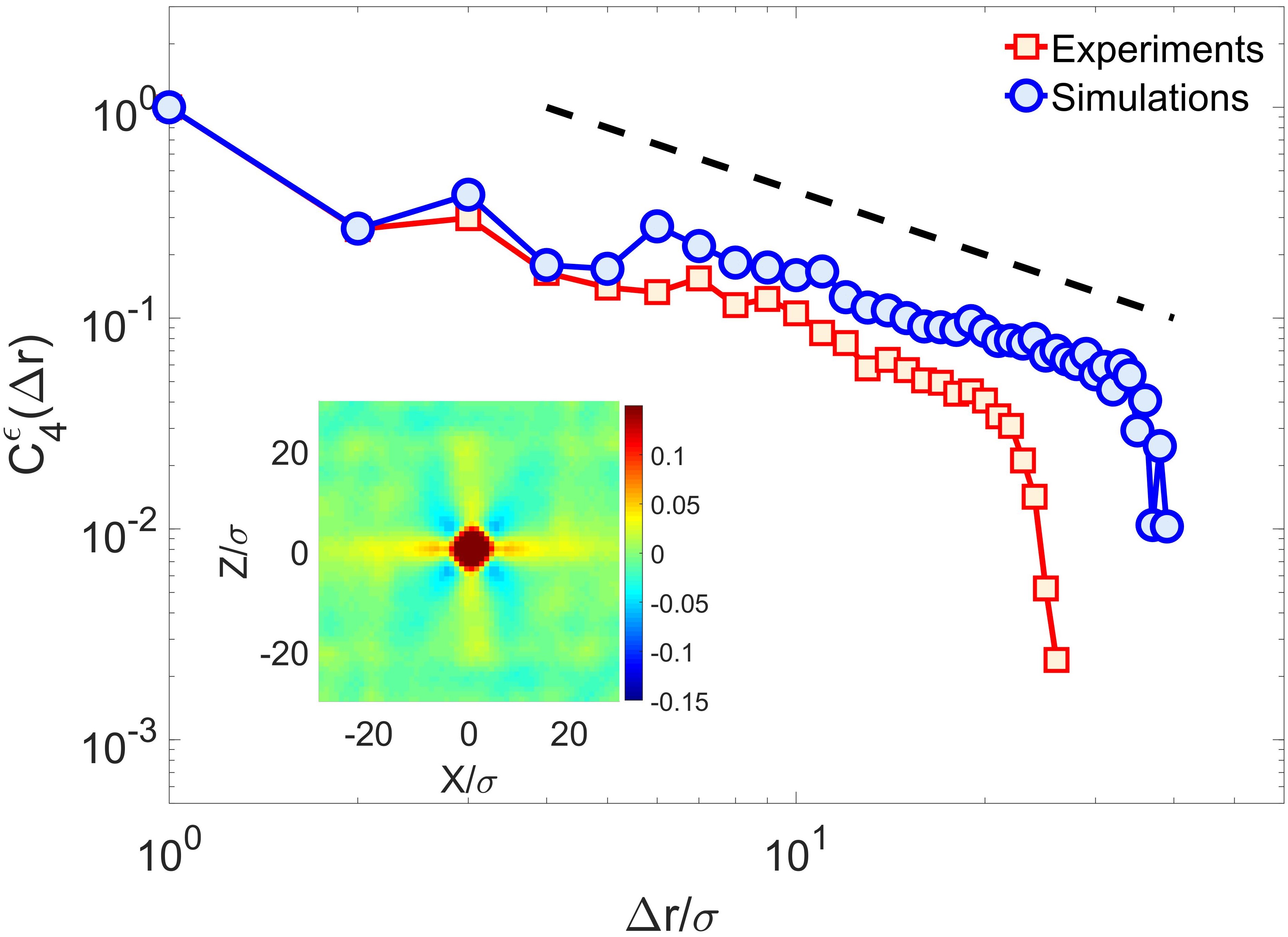}
\caption{Spatial correlations of microscopic strain averaged over all the particles in the steady state. Main panel : the projection of strain correlations from experiments is shown using square symbols and the simulation results are presented using circles. The dashed line shows a $1/\Delta r$ variation. Inset: the spatial correlations of $\epsilon_{xz}$ in the shear plane $xz$ obtained from the experiments. This corresponds to the correlations in Eq.~1 when $\Delta y=0$. The quadrupolar symmetry of strain correlations is evident.}
\label{fig1}
\end{figure}

We first present the analysis of microscopic strain and their spatial correlations in the steady state in experiments. The system is sheared at constant rate of $\dot{\gamma}\sim1.5\times10^{-5}s^{-1}$, and the flow is homogeneous in the steady state \cite{Chikkadi11}. A strain interval of $\Delta\gamma\sim0.036$ is used to compute individual particles' displacements and microscopic strain tensor. A reconstruction of the particle shear strain $\epsilon_{xz}$ is shown in Fig.~S1(a). The particles are color-coded based on $\epsilon_{xz}$; the red color indicates deformation in the direction of shear, and the blue color indicates deformation in the opposite direction. The particle scale deformation is heterogeneous with a network of positive and negative strain regions. The spatial correlations averaged over several strain steps of $\Delta \gamma\sim10^{-3}$ in the steady is shown in the inset of Fig.~1. This inset depicts the correlation in the shear plane, which is obtained when ${\bf \Delta r} = (\Delta x,0,\Delta z)$ in Eq.~1. The correlation has a familiar quadrupolar symmetry characteristic of the strain field around an Eshelby inclusion in homogeneous elastic materials. It suggests that high-strain regions (blue and red zones in Fig.~S1(a)) act as strained inclusions in an elastic matrix. Further, we project the correlation function onto its corresponding circular harmonic to determine radial correlations of strain following the expression : 
\begin{equation}
C_{\epsilon}^{4}(\Delta r)=\int_{0}^{2\pi} C_{\epsilon}(\Delta x,0,\Delta z)~cos(4\theta)~d\theta. 
\end{equation}
The $C_{\epsilon}^{4}(\delta r)$ curve with red squares in the main panel of Fig.~1 shows the projected radial correlations $C_{\epsilon}^{4}(\Delta r)$ corresponding to the polar plot in the inset. The dashed line has a slope unity and it is drawn for comparison. Similar to non-affine displacements \cite{Chikkadi11}, the strain correlations display an algebraic decay $1/r^{\alpha}$, where $\alpha\sim1$. However, this decay is slower than the $1/r^{3}$ variation predicted by Eshelby for a strained spherical inclusion in isotropic elastic solid \cite{Eshelby57}. The line with circles in Fig.~1 denotes the strain correlations obtained from finite shear rate simulations performed at $\dot{\gamma}=10^{-4}$. A good agreement between the curves confirms the robustness of strain correlations in two disparate systems. It appears that the $1/r$ decay stems from the strain correlations that capture the elastoplastic response of not a single inclusion but several interacting inclusions. This is also evident from the heterogeneous strain map in Fig.~S1(a), which points to the organized formation of multiple inclusions. 

\begin{figure*}[!]
\centering
\begin{tabular}{llll}
\includegraphics[width=.28\textwidth]{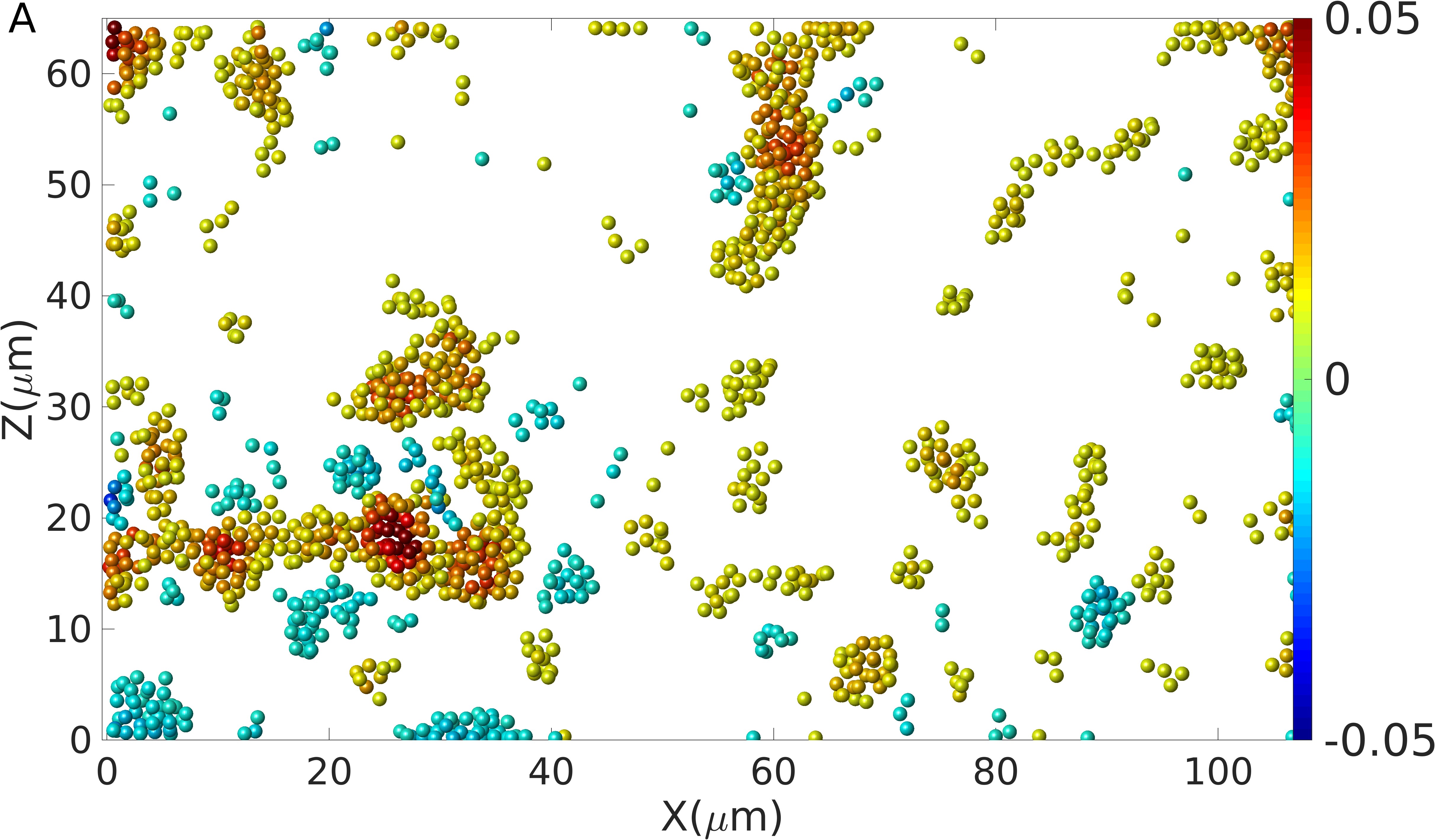}&
\includegraphics[width=.28\textwidth]{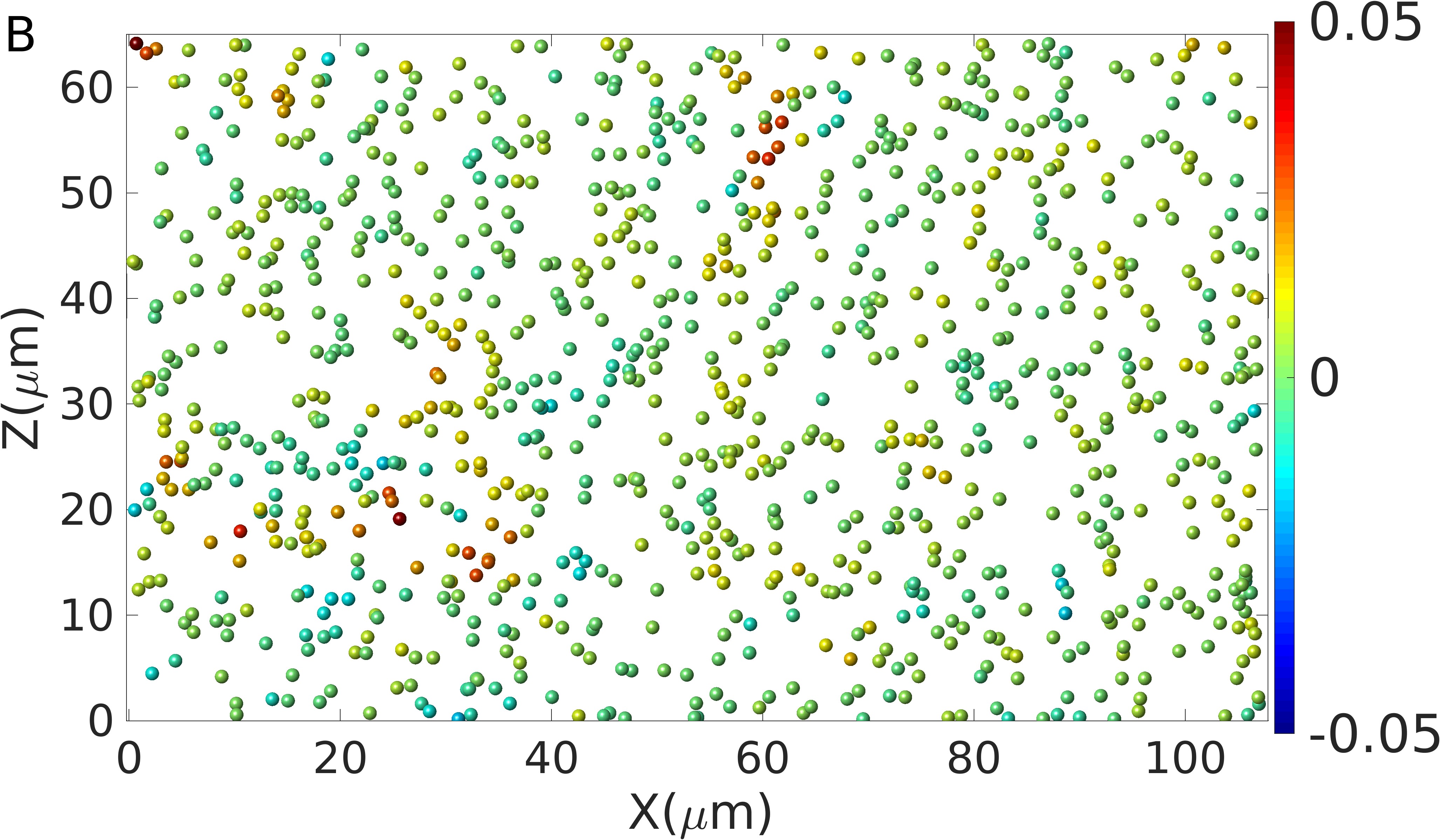}&
\includegraphics[width=.17\textwidth]{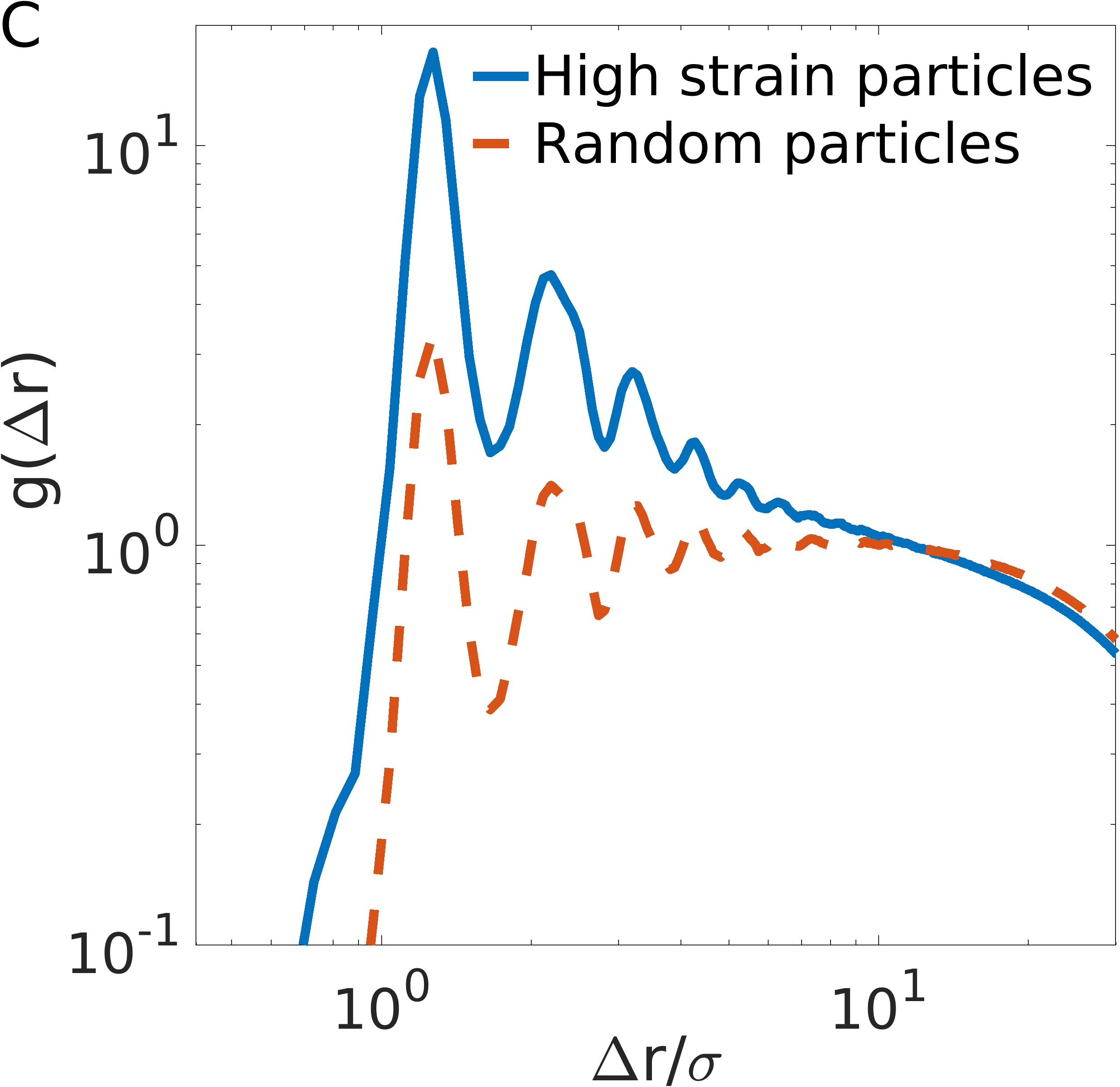}& 
\includegraphics[width=.17\textwidth]{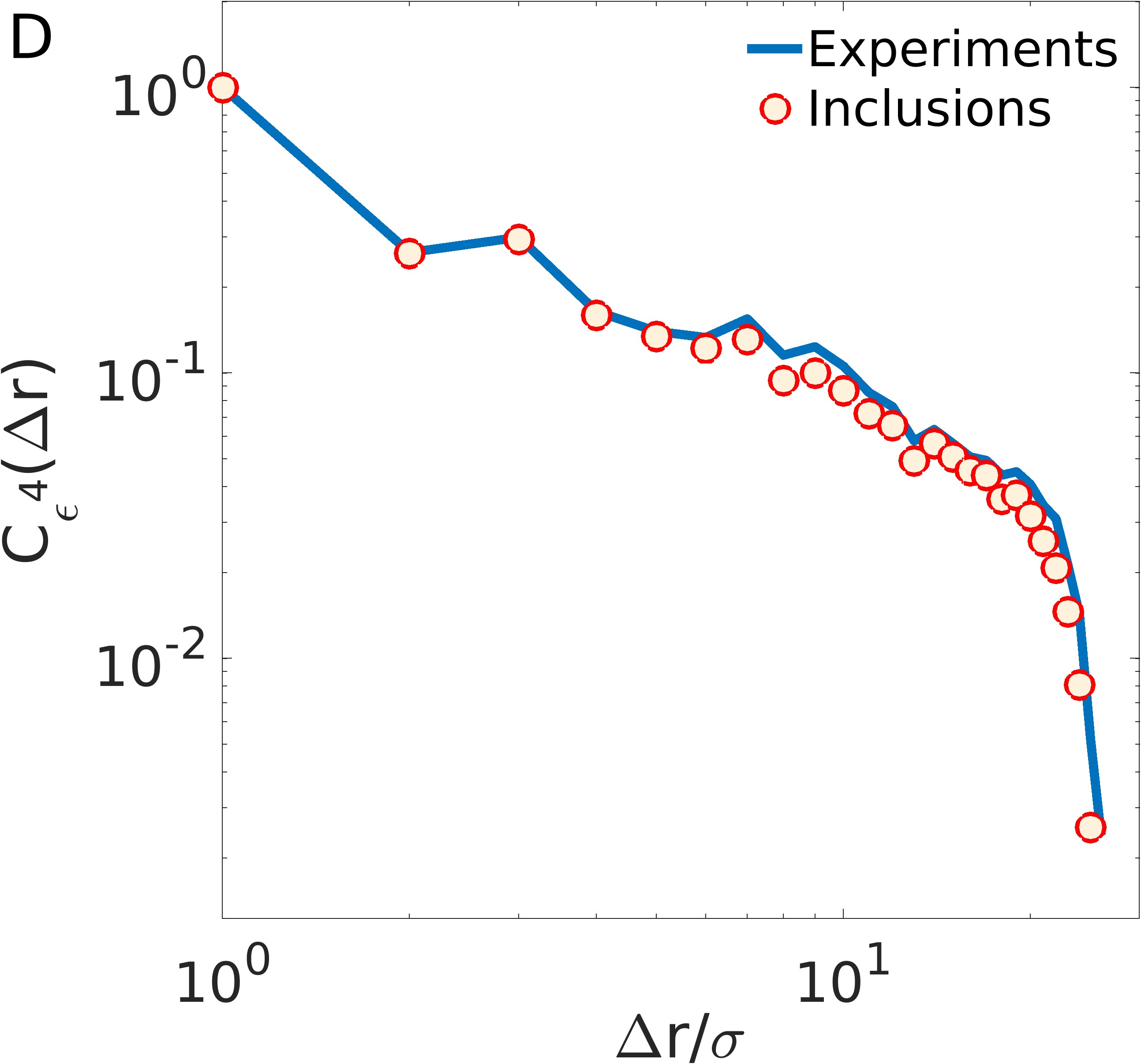}\\
\end{tabular}
\caption{(a) Spatial maps of particles with shear strain $\epsilon_{xz}>0.95\epsilon_{xz}^{max}$, where $\epsilon_{xz}^{max}$ is the maximum strain of particles. These are considered as inclusion centers in the first case. (b) Inclusion centers are obtained from a random selection of particles. Note that the particles in (a) and (b) are again shown in a $5\mu m$ thick region in $y-$direction. (c) The normalized pair correlation function of inclusion centers. The continuous line corresponds to inclusion centers obtained from particles with top $5\%$ strain, and the dashed line is obtained for the random selection of inclusion centers.}
\label{fig2}
\end{figure*}

Several investigations of amorphous plasticity and mesoscopic simulations have used a minimal model based on interacting Eshelby inclusions to understand elastoplastic features of sheared amorphous solids \cite{Dahmen09, Zoe17, Rodney11, Ratul12, Ratul13-1, Ratul13-2, Zaccone17}. Motivated by these ideas, we identify the inclusion centers in our experiments using microscopic strain and use them to construct synthetic strain fields. All the particles with shear strain exceeding a threshold value of $|\epsilon^i_{xz}|>\gamma_c$ are considered to be inclusions. This corresponds to the top $\sim10\%$ of particles in our system. It was verified that the results presented here are robust and not sensitive to the strain threshold. A spatial map of the inclusions thus identified is shown in Fig.~2(a). 
Apparently, they form clusters. To examine the effect of clustering, we also consider equal number of random particles as inclusions for comparison, which is shown in Fig.~2(b). The pair correlation function $g(r)$ of the inclusions centers is presented in Fig~.2(c). It is averaged over all inclusions in the system and several strain steps of $\Delta\gamma$ in the steady state. It is evident that the structural correlations are stronger when particles with top strains are considered as inclusion centers. In contrast, the inclusion centers corresponding to random selection are weakly correlated.

Next, we calculate the synthetic strain of particles due to the inclusions. The far-field shear strain around a single inclusion in an isotropic homogeneous elastic solid is $\varepsilon_{xz}^{I} \propto \frac{\epsilon_0 cos(4\theta )}{r^3}$ \cite{Barrat18}, where $\epsilon_0$ is the core strain, $\theta=cos^{-1}(z,r)$ is the azimuthal angle in the plane of shear and the particle diameter is core size. The core strain $\epsilon_0$ of an inclusion is assumed to be the shear strain of the particle obtained from experiments. The shear strain on any other particle in the system is computed as a superposition of shear strains due to all inclusions 
\begin{equation}
\epsilon_{xy}^{i}=\sum_{j=1}^{N_{inc}}\varepsilon_{xz}^{I,j}(\Delta r_{ij},\theta_{ij}),
\end{equation}
where $\Delta r_{ij}$ is the distance of particle $i$ from an inclusion $j$, and $\theta_{ij}= cos^{-1}(z_{ij},r_{ij})$. The reconstructions of shear strain when the inclusions are selected based on high-strain particles and randomly chosen are shown in the left column of Fig.~S2. It is not surprising that the deformation map in Figs.~S2(top left) compares well with experiments in Fig.~S1. To test our approach further, we have computed spatial correlations of synthetic strain by averaging over several strain steps, and these results are shown in the right column of Figs.~S2 (SI). Surprisingly, the strain correlations corresponding to high-strain inclusion centers display a quadrupolar symmetry in the top right panel of Fig.~S2. On the contrary, the symmetry is absent from the one corresponding to random inclusions in the bottom right panel of Fig.~S2. In the next step, the radial correlations for high-strain inclusions are computed and are shown together with experimental results in Figs.~2(d). There is an excellent agreement. These results establish that strain correlation for sheared systems can be modeled as an interacting system of Eshelby inclusions.

\begin{figure}[!]
\centering
\includegraphics[width=.4\textwidth]{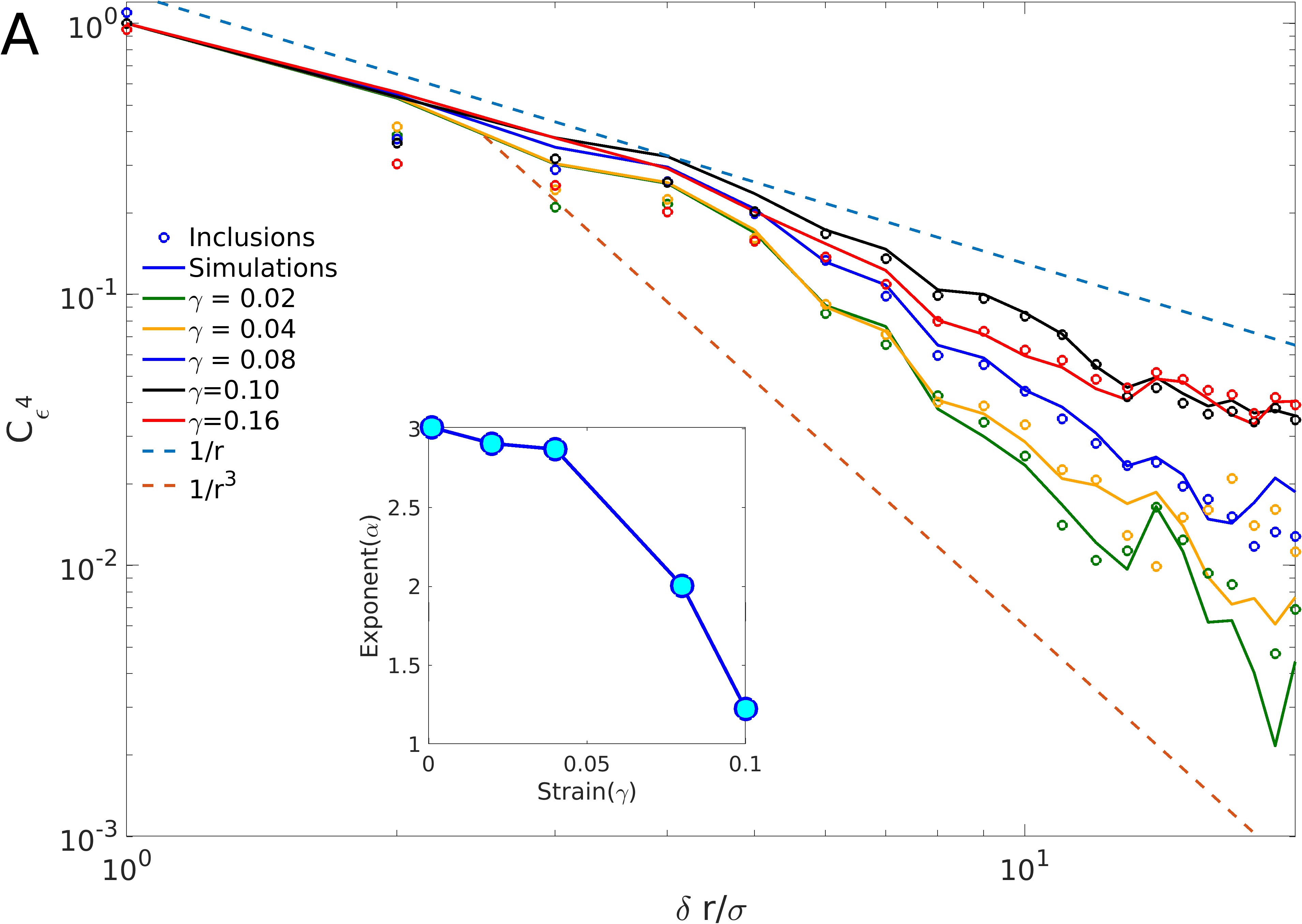} 
\includegraphics[width=.20\textwidth]{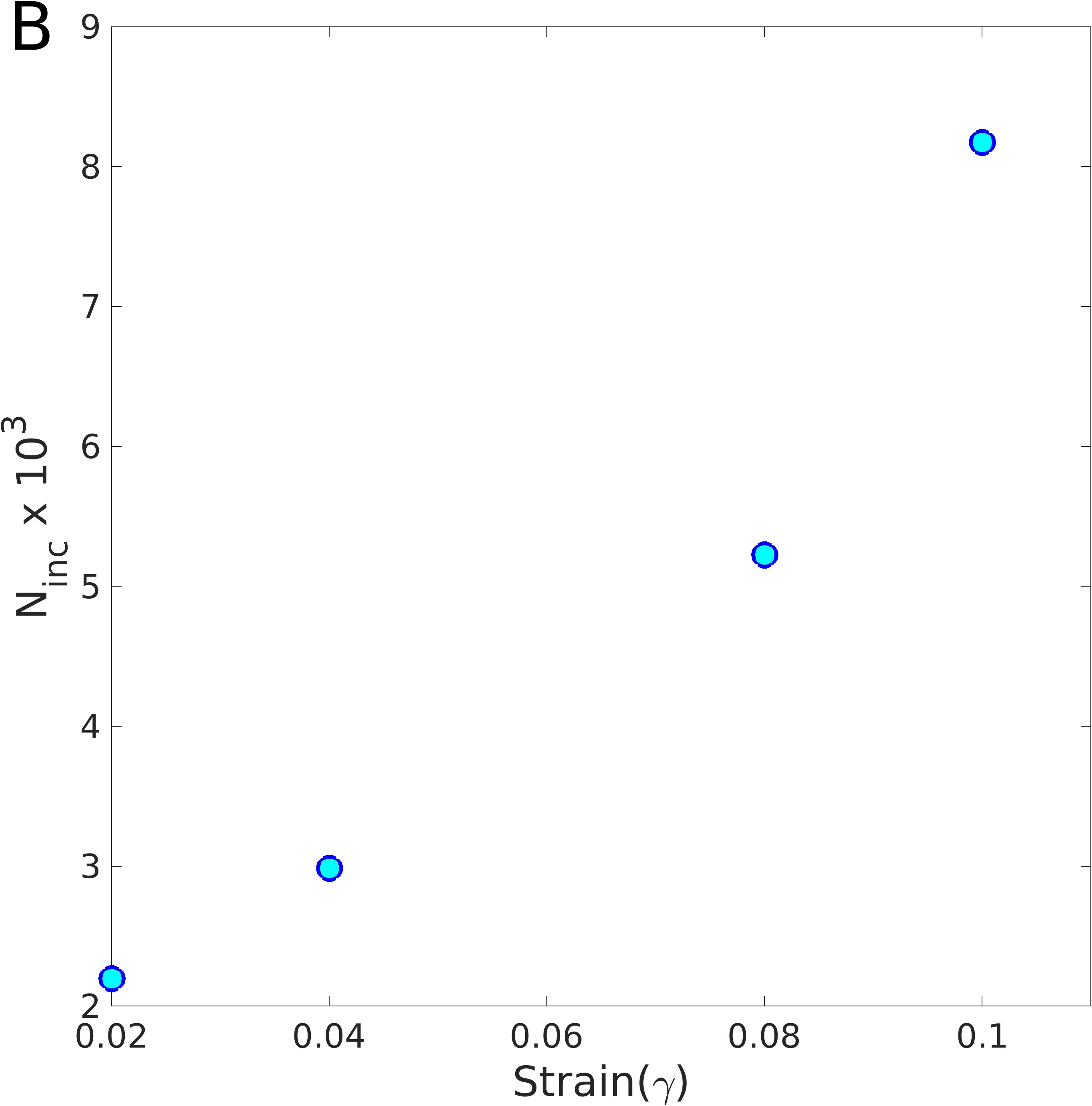} 
\includegraphics[width=.20\textwidth]{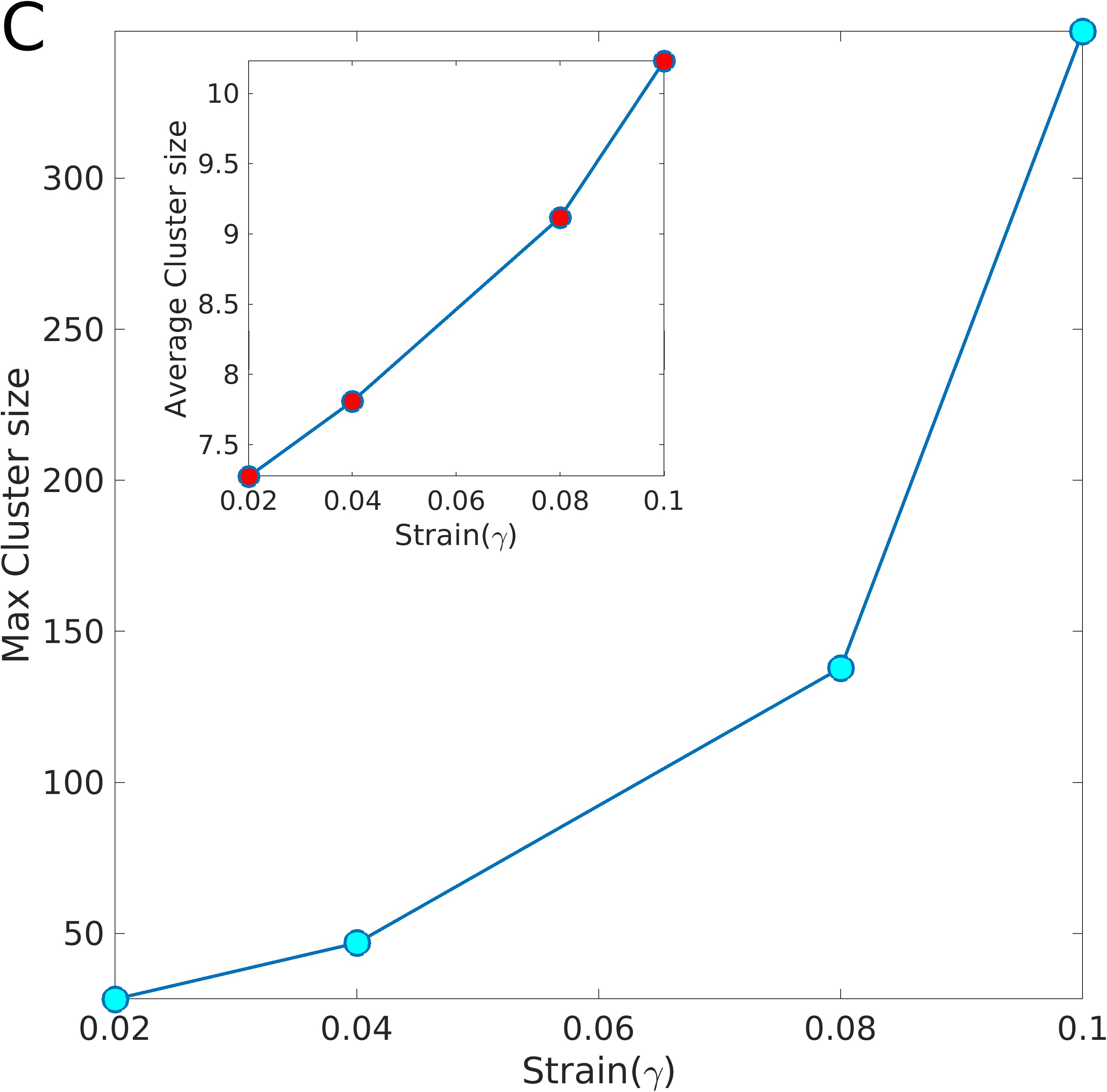} 
\caption{Inclusion centers and strain correlations in the transient stages of deformation. 
Top panel: The correlations of shear strain ($C^4_{\epsilon}(\Delta r)$) at various stages of deformation before the system attains a steady state is shown in the main panel. The thick lines are the simulation results. The symbols show the results obtained from inclusion analysis. The inset shows the variation in the value of exponent $\alpha$ as the system is strained. The exponent $\alpha$ is obtained from the best fits of correlations functions in the main panel.
Bottom left: The number of inclusions in the system as it is sheared from the linear elastic limit to steady state plastic flow. Bottom right: The inclusion centers form clusters. The size of the largest cluster is shown as a function of strain and the inset shows average cluster size. The results are averaged over $100$ realisations. 
}
\label{fig3}
\end{figure}

We next turn our attention to strain correlations in the transient stages of deformation before the system attains a steady state. The correlations in the steady state are averaged over several strain intervals. However, a similar averaging in the transient stages is not feasible in experiments due to a lack of sufficient realizations. Therefore, we exploit simulation data to investigate the transient stages. The strain correlations are computed using Eq.~1 at $\gamma=0.02, ~0.05, ~\text{and} ~0.1$ and are averaged over the azimuthal angle $\theta$ using Eq.~2. The results are shown in Fig.~3a using thick lines. The dotted lines with slopes $-1$ and $-3$ are shown for clarity. Surprisingly, the correlations decay with different exponents $1/r^{\alpha}$ as the system is sheared. The exponents obtained from data fitting are shown in the inset of Fig.~3a. When the strain is small, the exponent is $\alpha\sim3$. It varies continuously with increasing strain until $\alpha\sim1$ in the steady state. We elucidate the underlying physics by analyzing the synthetic strain fields. The first step is the identification of inclusions based on the threshold value $\left|\epsilon^i_{xz}\right|>\gamma_c$. The threshold value is set such that the number of inclusions in the steady state is $\sim 10\%$ of particles in the system. The Fig.~3b shows the count of inclusions with increasing strain. Apparently, the population of the inclusions grows monotonically as the system is sheared. A similar rise in the number of inclusions was reported by an earlier numerical study \cite{Rodney20}. A reconstruction of inclusion centers is shown in Fig.~S3(a)-(d) at $\gamma=~0.02, ~0.04, ~0.08, \text{and} ~0.1$, respectively. It is clear that the inclusions begin to cluster and grow as the system is sheared to a steady state. 
We delineate these changes in cluster sizes from cluster size distribution. The clusters are identified by setting a distance criterion of $r_c\leq \sigma_{AA}$ to identify pairs of particles as neighbors. The results of the analysis are shown in Fig.~3c. The size of the largest cluster in the main panel and the average size of the cluster in the inset display a growing trend with increasing strain. It establishes the observations made in Fig.~S3(a)-(d). To explain the changing value of the exponent $\alpha$ in Fig.3c, we compute synthetic strain fields from the inclusions. Again, all other particles' synthetic strain is obtained from the superposition of strain due to all inclusions. Further, their angle-averaged spatial correlations are determined from Eqs.~1 and 2. The result of these calculations is shown in the main panel of Fig.~3a. The continuous lines indicate the strain correlations obtained from simulations, and the symbols are obtained from the synthetic strain fields due to inclusions. The agreement between simulations and inclusion analysis is excellent. These results firmly establish that high-strain particles in the systems act as inclusion centers. A minimal model of interacting inclusions captures the strain correlations well in the transient and steady state of deformation of amorphous solids.

 In summary, we have investigated the elastoplastic deformation of amorphous solids by studying the spatial correlations of microscopic strain in experiments and simulations. The correlations show a familiar quadrupolar symmetry. However, their decay is anomalous $C_{\epsilon}^{4} (r) \sim 1/r^{\alpha}$, where $\alpha \sim 1$ in the steady state and $1<\alpha<3$ in the transient stages of deformation. This behavior is described well by a simple model of interacting inclusions. The inclusions are identified from the high-strain particles in the system. In the transient stages of deformation, when the strain on the system is small, the population of inclusion centers is low and weakly correlated. However, as the strain increases, the inclusions begin to grow and saturate in the steady state. Besides, they also form clusters, signaling a stronger spatial correlation. Due to this enhancement in density correlations, the correlations of synthetic strains show a gradual change in their form. The correlations $C_{\epsilon}^{4}(r) \sim 1/r^{\alpha}$, with the exponent taking values from $3$ in the linear regime to $1$ in the steady state. The outcomes of our experiments and simulations firmly establishes the universal nature of strain correlation in sheared amorphous materials that are independent of the interactions and microscopic scale of the system.

\section*{Acknowledgements}
V.C. acknowledges startup grant from IISER Pune. A.G acknowledges support from Department of Science and Technology (DST), India for a WOS grant no.SR/WOS-A/PM-34.

\end{document}


\title{Supplementary information \\
Microscopic strain correlations in sheared amorphous solids}
\author{Sagar Malik$^{1}$, Meenakshi L.$^{2}$, Atharva Pandit$^{1}$, Antina Ghosh$^{1}$, Peter Schall$^{3}$, Bhaskar Sengupta$^{2}$, Vijayakumar Chikkadi $^{1}$}
\affiliation{
$^1$ Physics Division, Indian Institute of Science Education and Research Pune, Pune 411008, India.\\
$^2$ Department of Physics, School of Advanced Sciences, Vellore Institute of Technology, 632014 Vellore, India.\\
$^3$ Institute of Physics, University of Amsterdam, Amsterdam, The Netherlands.\\
}

\maketitle

\section{Materials and methods}
\subsection{Experimental methods}
We prepared a colloidal glass by suspending sterically stabilized fluorescent polymethylmethacrylate particles in a density and refractive index matching mixture of cycloheptyl bromide and cis-decalin. The particles have a diameter of $\sigma = 1.3 \mu m$, and a polydisperity of $7\%$ to prevent crystallization. The suspension was centrifuged at an elevated temperature to obtain a dense sediment, which was subsequently diluted to get a suspension of desired volume fraction $\phi\sim0.60$. The sample was sheared using a shear cell that had two parallel boundaries separated by a distance of $\sim 50\sigma$ along the $z-$direction. A piezoelectric device was used to move the top boundary in the $x-$direction to apply shear rates in the range $10^{-5}-10^{-4}s^{-1}$. To prevent boundary-induced crystallization in our samples, the boundaries were coated with a layer of polydisperse particles. Confocal microscopy was used to image the individual particles and to determine their positions in three dimensions with an accuracy of $0.03 \mu m$ in the horizontal and $0.05 \mu m$ in the vertical direction. We tracked the motion of $\sim 2 \times 10^{5}$ particles during a $25$-min time interval by acquiring image stacks every $60~s$. All the measurements presented here were made in the steady state, after the sample had been strained to $100\%$ at shear rate of $1.5\times 10^{-5}s^{-1}$, as confirmed by other independent rheological measurements.

\subsection{Simulation methods}
To prepare the thermal glass we use the well studied Kob-Anderson binary mixture model \cite{Kob}. Our model consists of $N$ classical point particles confined in a three-dimensional simulation box in the $NVT$ ensemble.  We choose a binary Lennard-Jones (LJ) mixture of particles which are labeled as A and B, and their number ratio is $80:20$. For simplicity, the mass of both types of particles $m$ is taken to be the same and equal to unity. The interaction potential for a pair of particles has the following form,
\begin{eqnarray}
U_{\alpha\beta}(r) &=& 4\epsilon_{\alpha\beta}\Big[\Big(\frac{\sigma_{\alpha\beta}}{r}\Big)^{12} - \Big(\frac{\sigma_{\alpha\beta}}{r}\Big)^{6} + A_0 + A_1\Big(\frac{r}{\sigma_{\alpha\beta}}\Big) + A_2\Big(\frac{r}{\sigma_{\alpha\beta}}\Big)^2\Big] \ , r\le r_{cut} \nonumber \\
&=& 0, \,~~ {r\textgreater r_{cut}}, \label{Uij}
\end{eqnarray}
where $\alpha, \beta \in  \rm{A, B}$. The units of various quantities in our simulation are as follows: lengths are expressed in the unit of $\sigma_{AB}$, energies in the unit of $\epsilon_{AB}$, time in the unit of $(m\sigma_{AB}^2/\epsilon_{AB})^{1/2}$ and temperature in the unit of $\epsilon_{AB}/k_{\rm B}$. Here, $k_{\rm B}$ is the Boltzmann constant which is unity. The parameters $\sigma_{\alpha\beta}$ and $\epsilon_{\alpha\beta}$ are chosen as follows: $\sigma_{AA} = 1.0, \sigma_{BB} = 0.88, \sigma_{AB} = 0.8$ and $\epsilon_{AA} = 1.0, \epsilon_{BB} = 0.5, \epsilon_{AB} = 1.5$. With these set of parameters, a binary LJ mixture avoids crystallization below $T_g$ and forms stable glass. $T_g$ for this model is approximately $0.44$ in the reduced unit \cite{Kob}. To improve the computational efficiency, the potential is truncated at $r_{cut}=2.5\sigma_{AA}$. The parameters $A_0 = 0.040490$, $A_1 = -0.009702$ and $A_2 = 0.000620$ are such that the potential and its derivatives (up to second order) go to zero smoothly at $r=r_{cut}$. 

To prepare the glass samples, we use MD simulation. All simulations are carried out at density $\rho=1.2$ and a total number of particles $N=150000$. Position and velocity of particles are updated using the velocity-Verlet integration technique \cite{Verlet} with time step $\Delta t =0.005$. The temperature of the system is kept fixed to the desired value by employing the Berendsen thermostat \cite{Berendsen}. Also, we apply periodic boundary conditions in all directions. 

We begin our simulation by equilibrating the binary mixture in the liquid state at temperature $T=1.0$. The system is then cooled to the final temperature $T=0.3$ below the glass transition temperature $T_g=0.43$  with a slow cooling rate $10^{-5}$ to prepare well relaxed glassy samples. For statistical averaging, the whole process is repeated $100$ times with different initial realizations. Finally, the glassy samples are sheared along the $xz$ plane at a constant shear rate $\dot{\gamma}=10^{-4}$. During the shear process, the temperature is controlled by dissipative particle dynamics thermostat \cite{DPD}. 

\subsection{Calculation of strain tensor and spatial correlations}
To compute these quantities we follow particle trajectories and identify the nearest neighbors of each particle as those separated by less than $r_{0}$, the first minimum of the pair correlation function. We subsequently determine the best affine deformation tensor ${\bf\Gamma}$ describing the transformation of the nearest neighbor vectors, ${\bf d_i}$, over the strain interval $\Delta \gamma$ \cite{Falk98}, by minimizing $D^2_{min} = (1/n) {\sum_{i=1}^{n}}({\bf d_i}(\gamma + \Delta \gamma) - {\bf \Gamma}{\bf d_i}(\gamma))^2$. The symmetric part of ${\bf \Gamma}$ is the local strain tensor. The remaining non-affine component $D^2_{min}$ has been used as a measure of plastic deformation. We will investigate the correlations in the fluctuations of the principal shear strain component $\epsilon_{xz}$ using the following expression \cite{Chikkadi11}:
\begin{equation}
C_{\epsilon}({\bf \Delta r}) = \frac{ \left< \epsilon_{xz}({\bf r + \Delta r}) \epsilon({\bf r})
\right> - \left< \epsilon_{xz}({\bf r}) \right> ^{2} } { \left< \epsilon_{xz}({\bf r})^{2}
\right> - \left< \epsilon_{xz}({\bf r}) \right> ^{2} }  ,
\label{c_r}
\end{equation}
where angular brackets denote average over all the particles in the systems and several strain steps. $C_{\epsilon}$ correlates values of $\epsilon_{xz}$ at locations separated by ${\bf \Delta r}$, this way we capture the elastic response of the system in the steady state.

\section{Microscopic shear strain and spatial correlations}

\subsection{Results from experiments in the steady state}
\begin{figure}[h]
\centering
\includegraphics[width=.32\textwidth]{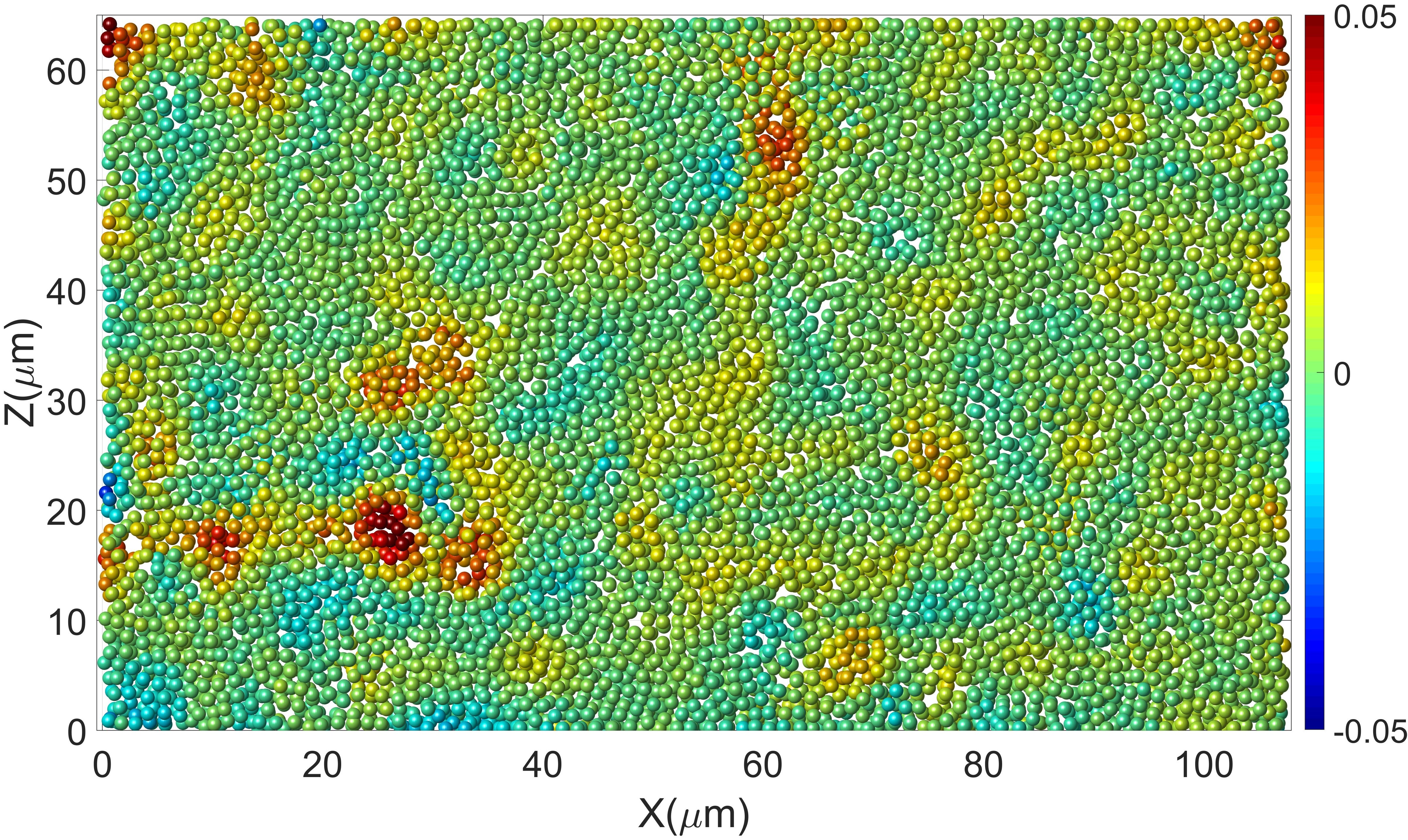}
\caption{Strain map in sheared colloidal glasses at a shear rate of $\dot{\gamma}\sim1.5\times10^{-5}s^{-1}$. The particles in a small section ($\sim3\sigma$) along y-direction are color coded based on their shear strain $\epsilon_{xz}$ values. Red color indicates deformation in direction of shear and blue color indicates deformation in the opposite direction.}
\end{figure}


\begin{figure}[h]
\centering
\begin{tabular}{cc}
\includegraphics[width=.3\textwidth]{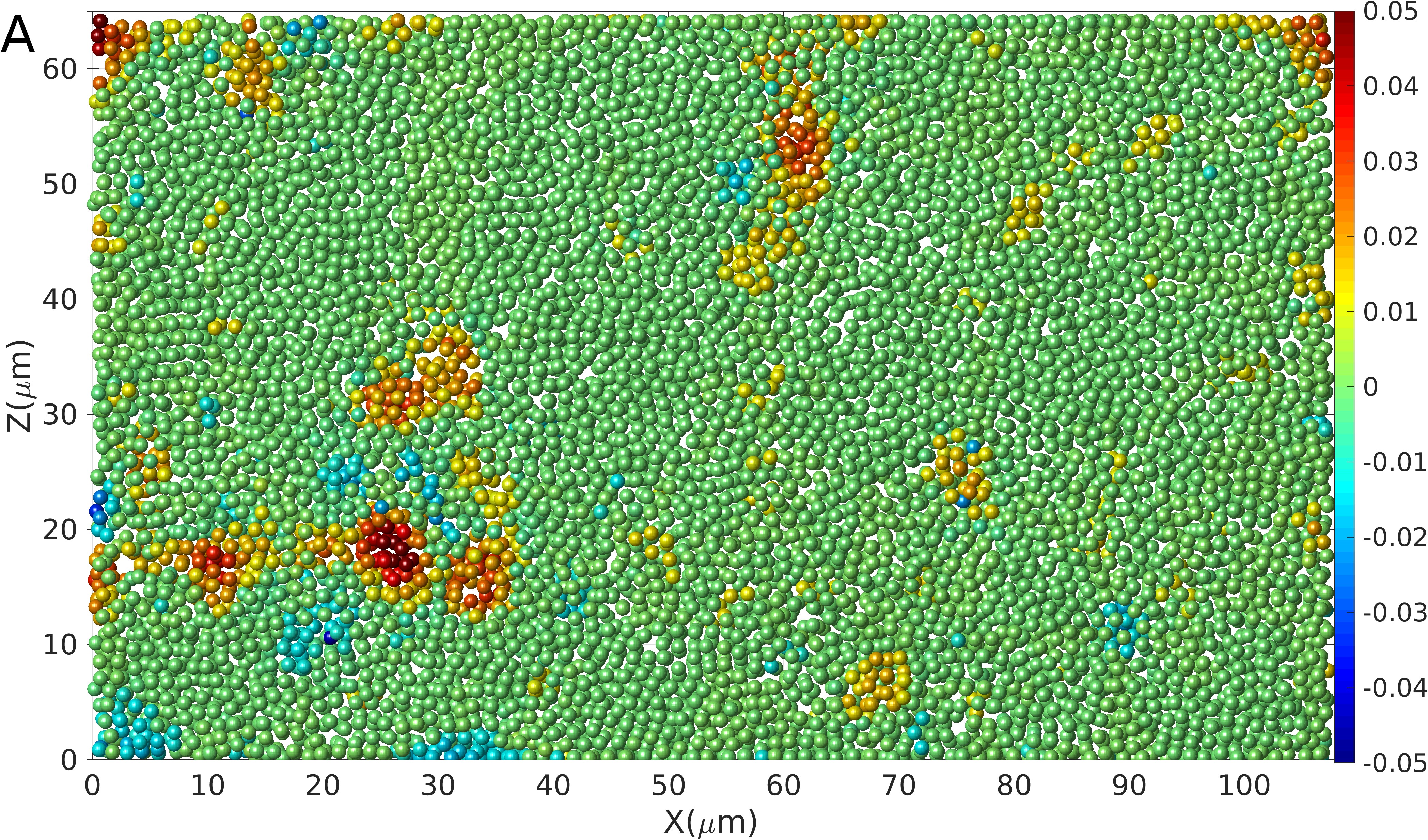} & 
\includegraphics[width=.2\textwidth]{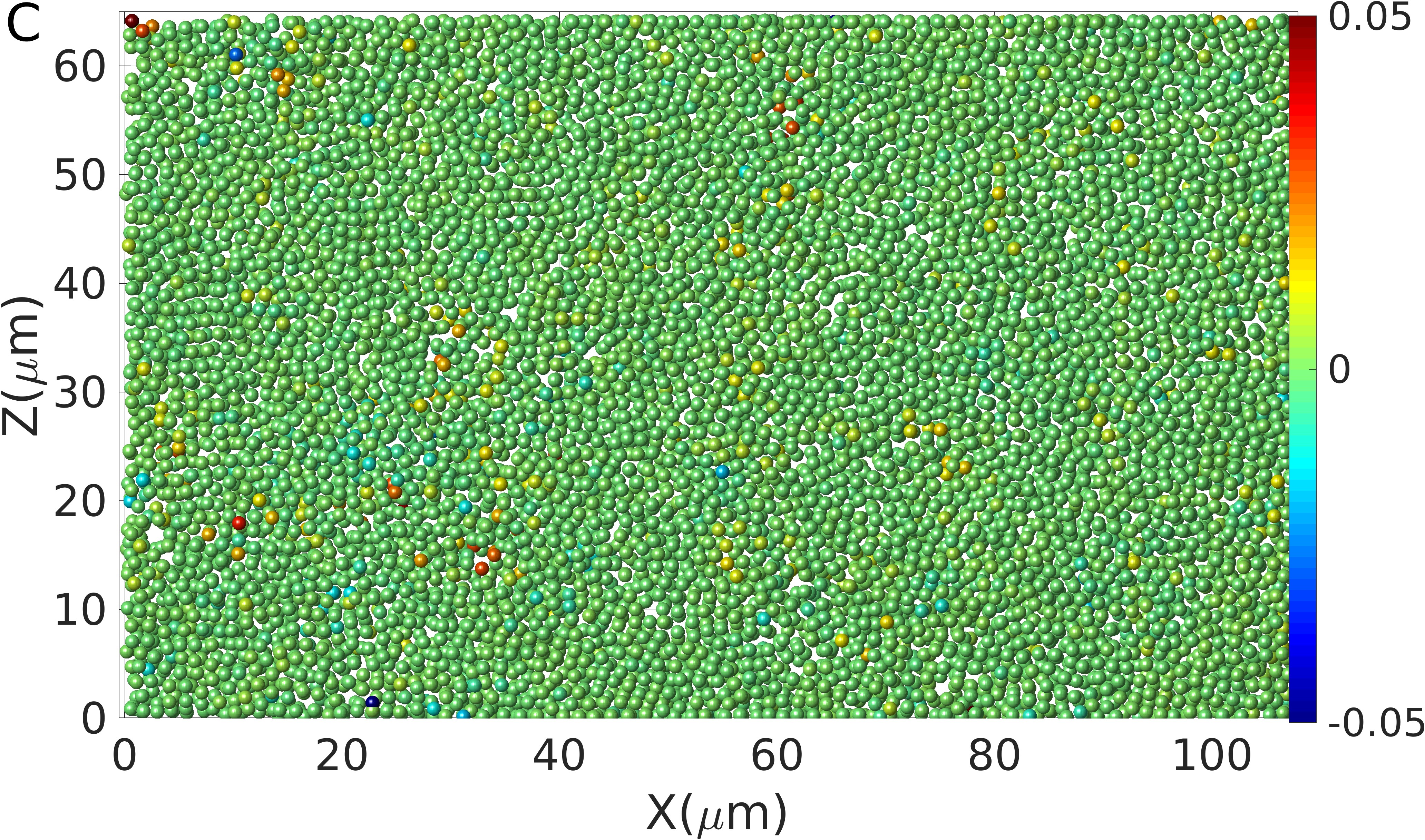} \\ 
\includegraphics[width=.3\textwidth]{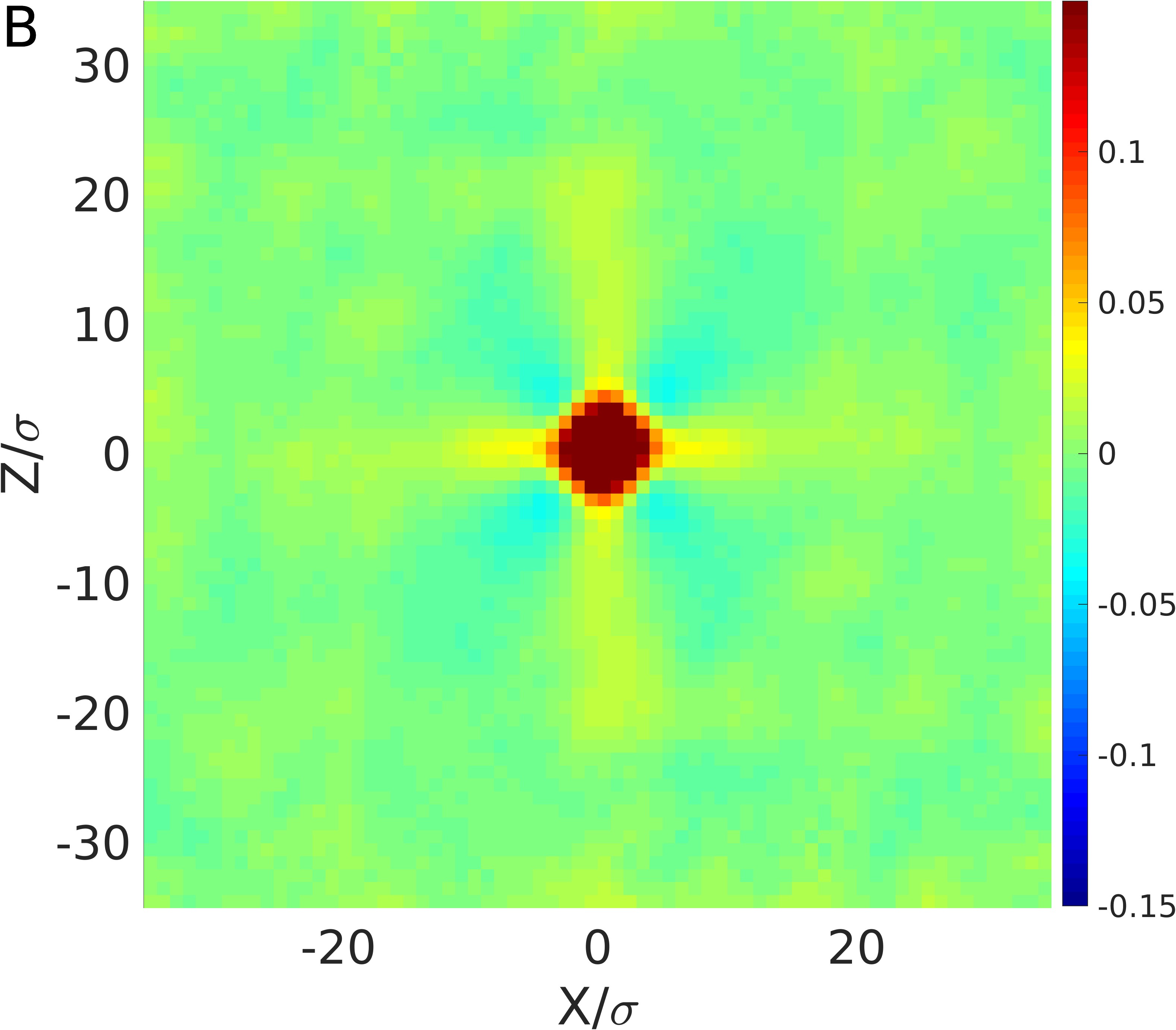} &
\includegraphics[width=.2\textwidth]{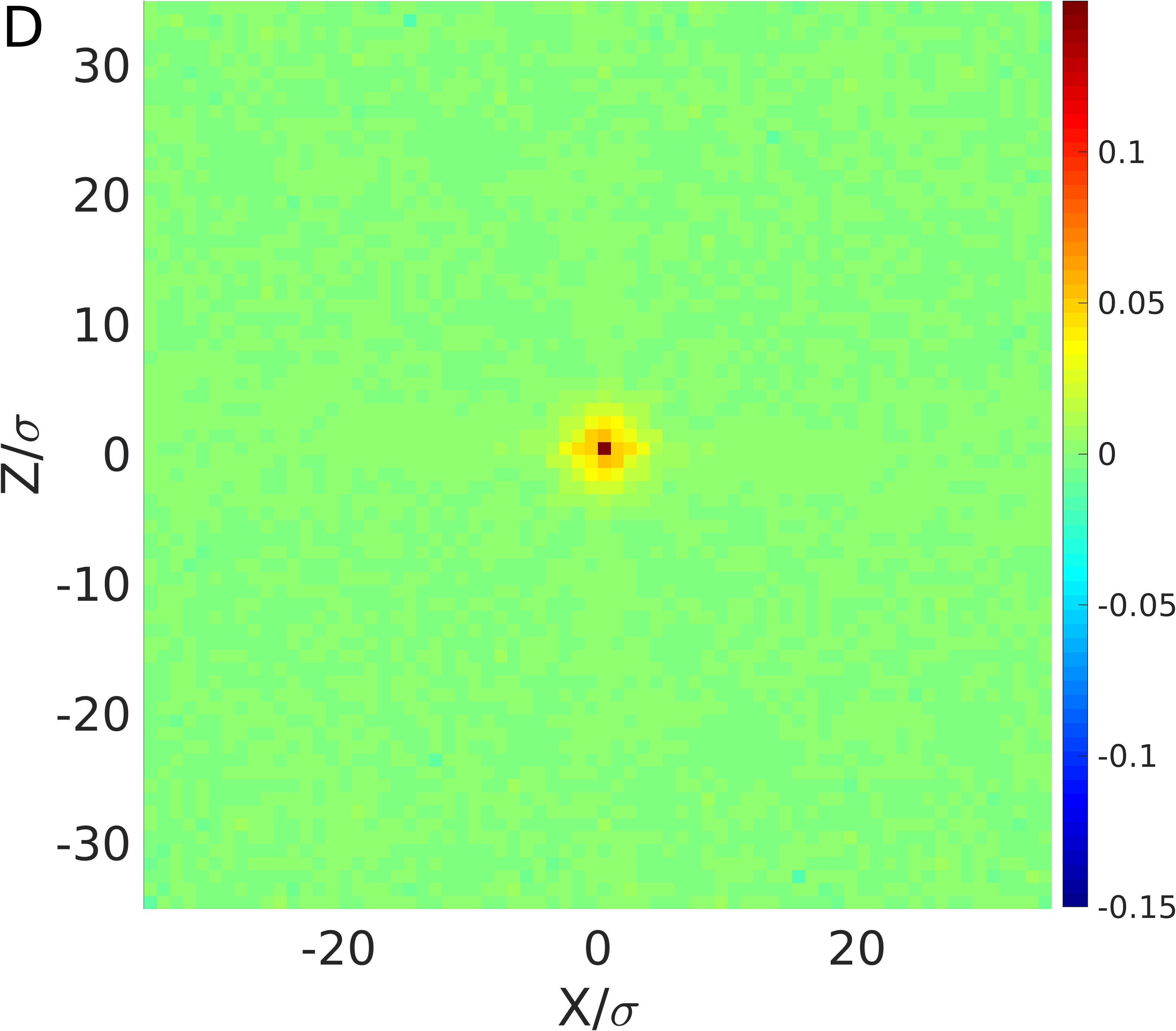} \\
\end{tabular}
\caption{The synthetic strain maps obtained from inclusions centers is shown in left panels and the right panels display their spatial correlations. The synthetic strain values are computed based on inclusions centers that were identified from high strain particles (top left) and selected randomly (bottom left). The spatial correlations of strain fluctuations when high strain particles are considered as inclusions (top right) and when inclusions are selected randomly (bottom right)}
\end{figure}

\subsection{Clustering of inclusions in the transient stages of deformation}

\begin{figure}[h]
\centering
\includegraphics[width=.22\textwidth]{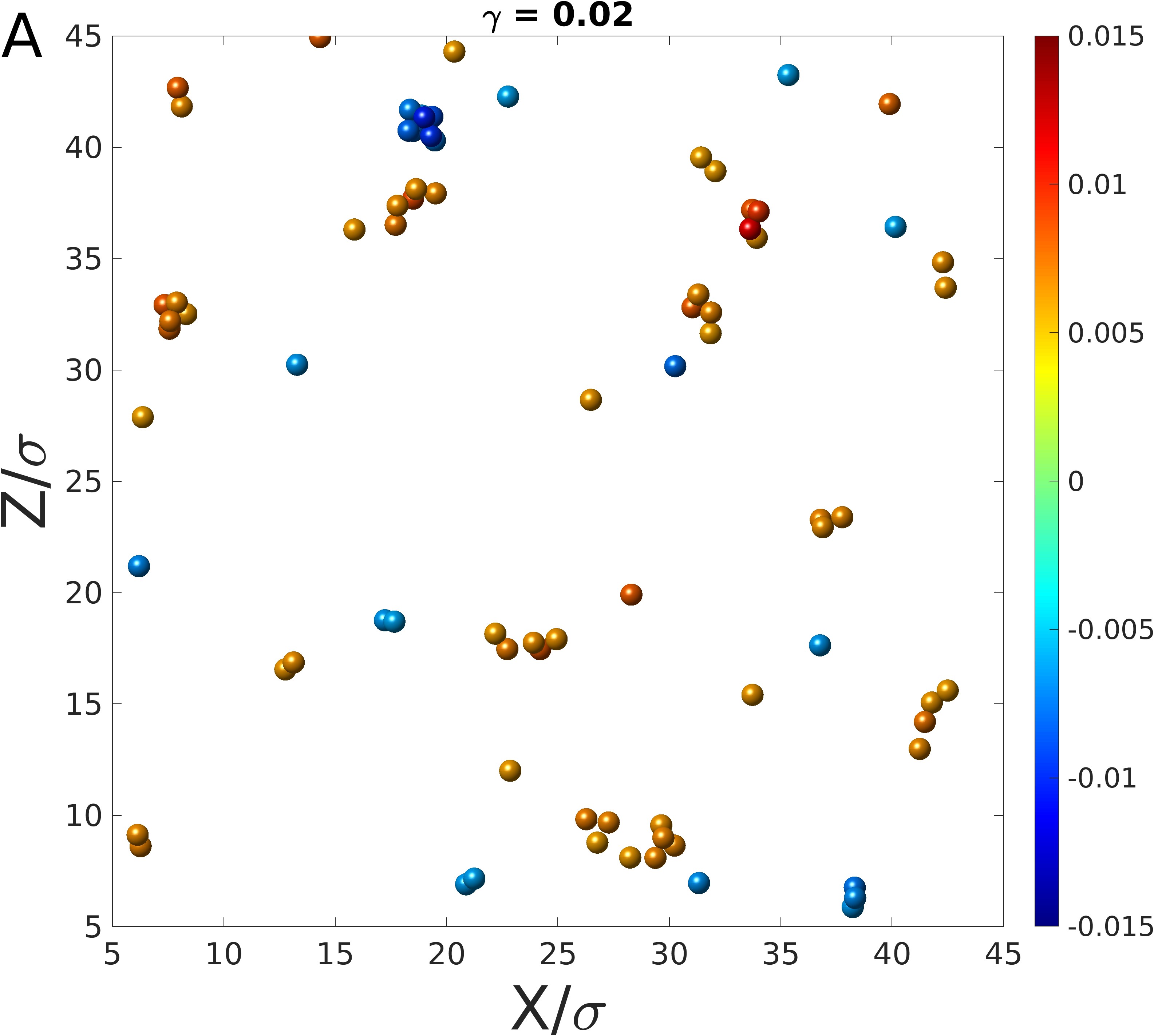}  
\includegraphics[width=.22\textwidth]{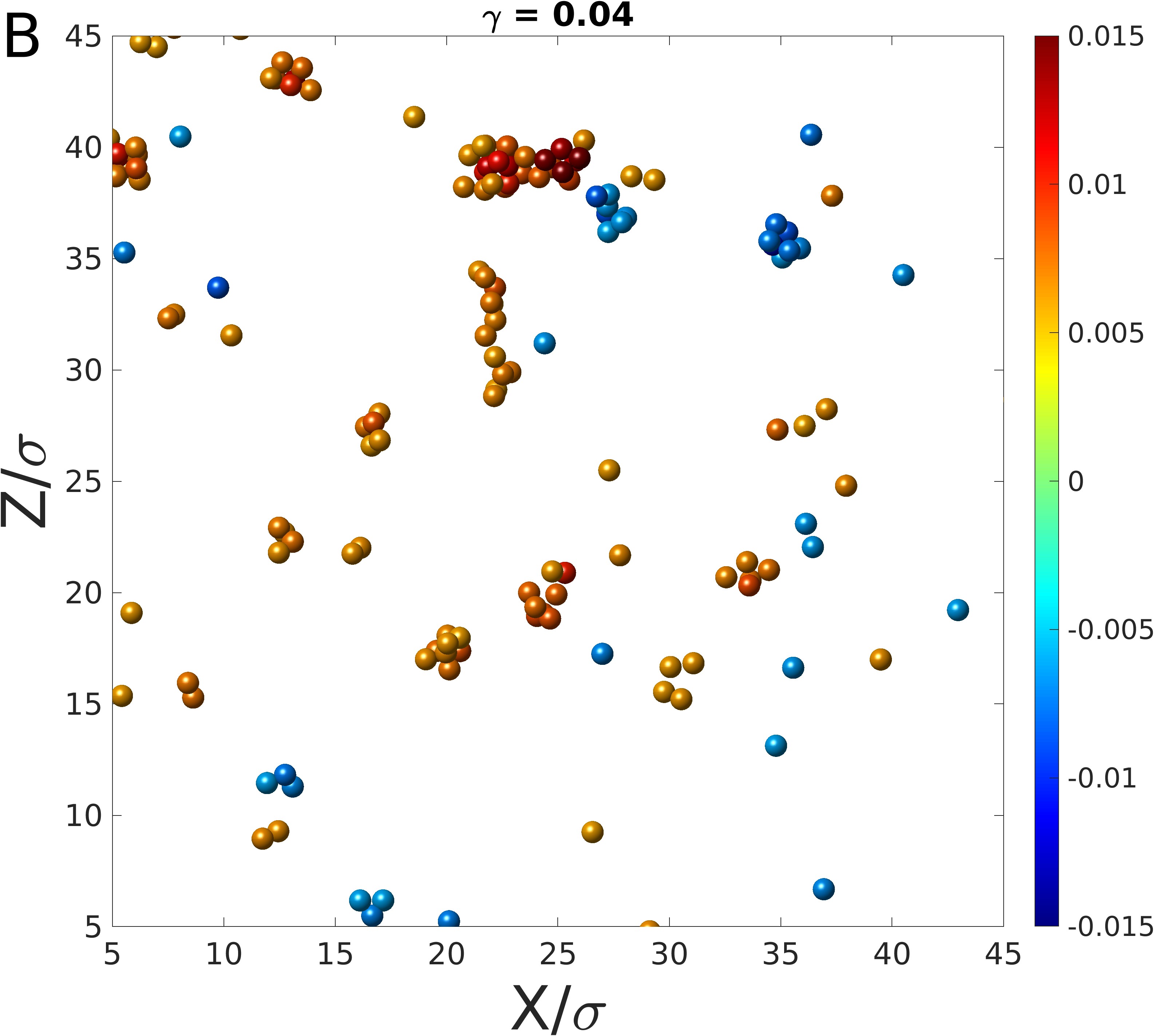} 
\includegraphics[width=.22\textwidth]{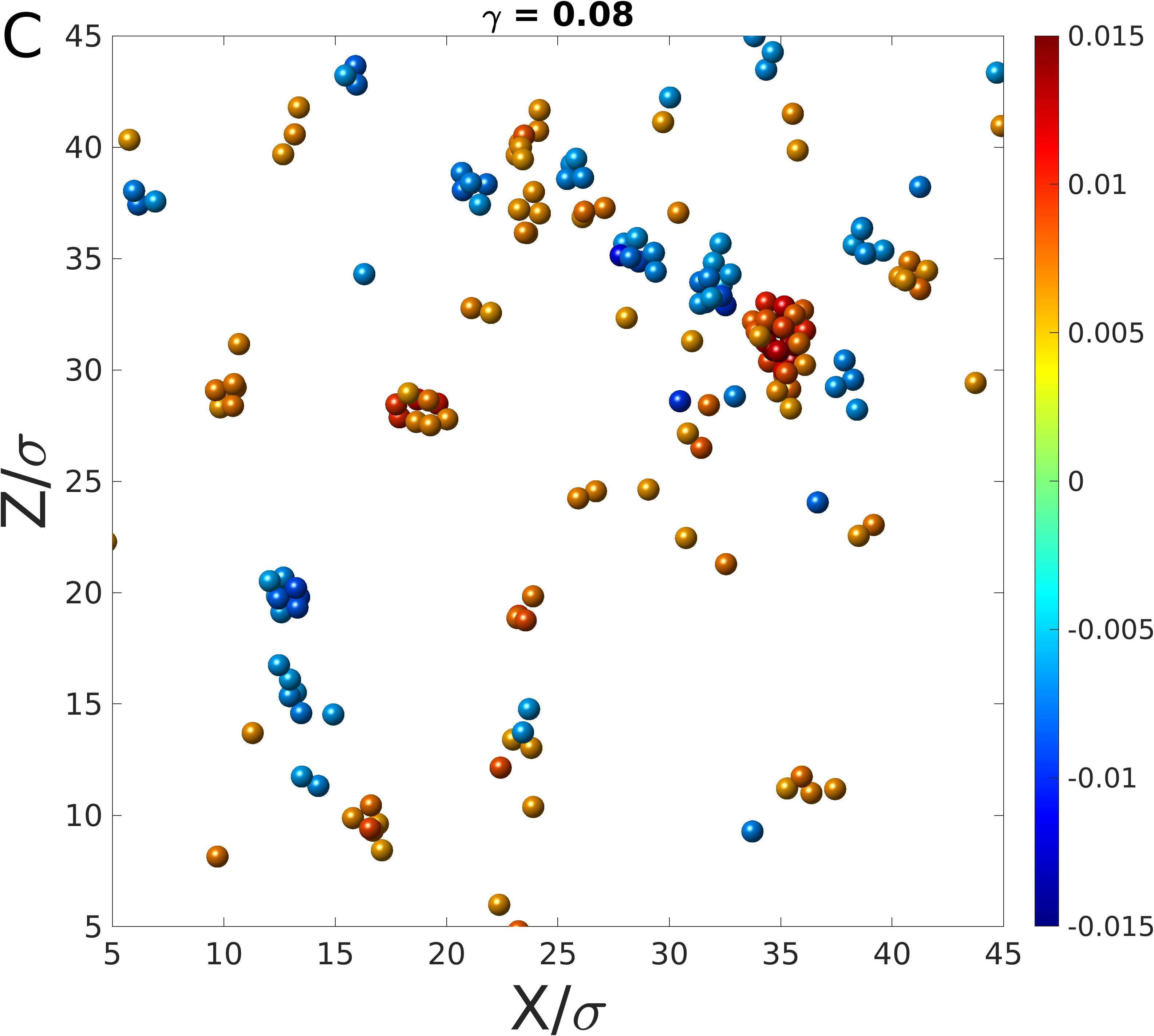} 
\includegraphics[width=.22\textwidth]{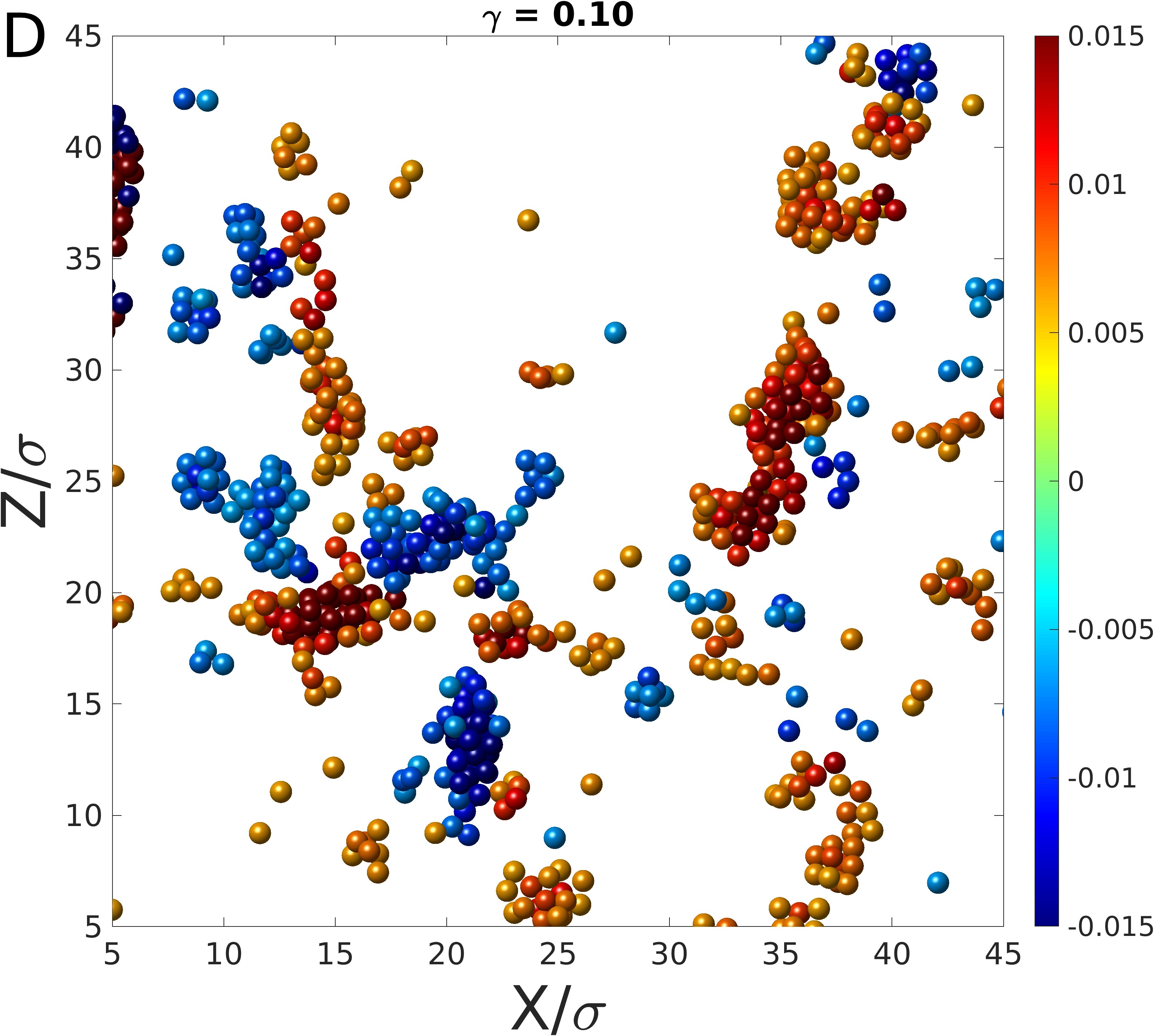}
\caption{A spatial map of inclusion centers in the transient stages of deformation. The inclusions are identified based on a threshold value of shear strain $\epsilon_i>\gamma_c$, where $\gamma_c$ is defined in the main text. The inclusions in a section of $3\sigma_{AA}$ thickness are shown at $\gamma=0.02$ (a), $\gamma=0.04$ (b), $\gamma=0.08$ (c)  and $\gamma=0.1$ (d). As the system is strained, the inclusions begin to cluster}
\end{figure}